\documentclass[twocolumn,showpacs,preprintnumbers,amsmath,amssymb,eps]{revtex4}
\usepackage{graphicx}
\usepackage{dcolumn}
\usepackage{bm}

\begin{document}

\title{Phase Separation and the Dual Nature of the Electronic Structure in Cuprates}
\author{E. V. L. de Mello}
\altaffiliation[]{evandro@if.uff.br}
\author{D. H. N. Dias }
\affiliation{
Instituto de F\'{\i}sica, Universidade Federal Fluminense, Niter\'oi, RJ 24210-340, Brazil\\}
\date{\today}
\begin{abstract}

The dual nature of the  electronic structure of stripes 
in $La_{2-x}Sr_xCuO_4$ was characterized by many experiments.
We present here an attempt 
to characterize this  dual behavior based on the Cahn-Hilliard theory
of a  phase separation transition  which is assumed to occur at
the high pseudogap temperature. The resulting inhomogeneous 
low doping system is formed  of hole-rich (metallic like) regions embedded in
a hole-poor (insulator like). This inhomogeneous configuration  
is analyzed by a new method within the Bogoliubov-deGennes  superconducting theory.
This approach describes well the electronic nodal-antinodal dichotomy and 
parts of the phase diagram.
%
%
\end{abstract} 


\maketitle 


The high critical temperature superconductors
(HTSC) represents today one of the greatest challenge of
condensed matter physics. It is likely that the main difficulty
to understand
the physical properties of HTSC is due to
the fact that some families seem to have
a high inhomogeneous electronic structure while others  families appear to
be more homogeneous or, at least, without some  gross inhomogeneity.

An important technique to study the HTSC electronic structure 
is provided by
angle resolved photoemission (ARPES) experiments. With the improvement of the
energy and momentum  resolution in  recent 
works\cite{DHS,Zhou,Zhou04,Ino,Yoshida}, it was possible
to distinguish a two component electronic structure in 
$\vec k$-space of the $La_{2-x}Sr_xCuO_4$ (LSCO) family: a metallic quasi particle
peak crossing the Fermi level along the zone
diagonal following the  ($0,0$)-($\pi,\pi$) nodal direction which
increases with the doping level\cite{DHS,Yoshida}. On the other hand, 
the spectral weight at the straight 
segments  in the ($\pi,0$) and (0,$\pi$) antinodal regions, is compatible
with a quasi one dimensional structure or with static 
stripes\cite{Tranquada}. Due to
the d-wave symmetry of the superconducting order parameter, the zero temperature
superconducting gap $\Delta_0$ vanishes at the nodal and 
is measured at the antinodal
directions by the leading edge shift on the Fermi
surface by the ARPES spectra. Moreover, the
values of $\Delta_0$  decreases with doping
but the quasiparticle spectral weight near the nodal 
directions, at the Fermi level,
increases, showing the distinct behavior of 
these two aspects of the electronic structure\cite{DHS,Zhou,Zhou04,Ino}.

A complementary technique to ARPES  is provided by 
scanning tunneling microscopy (STM) since it probes the differential
conductance or $\Delta_0$ directly on the surface of the compound.
Recent STM data  have revealed a patchwork of 
(nanoscale) local spatial variations in the density of states which is
used to measure the local superconducting gap\cite{Pan,Davis}.
With this technique, it was also possible to distinguish two distinct 
behavior: well defined coherent
and ill-defined incoherent peaks depending on the spectra location 
on a $Bi_2Sr2CaCu_2O_{8+\delta}$ (Bi2212) surface\cite{Hoffman,McElroy1,McElroy2}. 
Also, tunneling experiments using
superconductor insulator superconductor (SIS) with 
different insulator layers  have shown distinct sets of energy scales
and have also led to the idea that the richness of the phase diagram as 
function of doping is due to the charge inhomogeneity in the Cu-O 
planes\cite{Mourachkine}.

These and others unusual features of cuprates led to theoretical proposals that
phase separation is essential to understand their physics\cite{Zaanen,Emery}.
In fact, phase segregation has been observed on the $La_2CuO_{4+\delta}$ system by
x-ray and transport measurements\cite{Grenier,Jorg}. They have
measured a spinodal phase segregation into an oxygen-rich (and hole-rich)
metallic phase and an oxygen-poor antiferromagnetic phase above T=220K.
Below this temperature the mobility of the interstitial oxygen becomes
too low for a further segregation. $La_2CuO_{4+\delta}$  is the only 
system where ion diffusion has been firmly established, although there are
evidence of ion diffusion at room temperature in microcrystals of the 
Bi2212 superconductors at a very slow rate\cite{Truccato}.

 In this paper we  take the large pseudogap temperature of the
HTSC phase diagram\cite{Tallon,Mello04} as the phase separation
temperature $T_{ps}(p)$. To obtain quantitative results on the charge 
separation as a function of the temperature, we apply the
Cahn-Hilliard (CH) theory\cite{CH} to this transition.  It yields the phase 
separation patterns found in many HTSC and it provides also an interpretation 
to the energy $E_g$, common to all cuprates\cite{Tallon},
as the potential barrier between the two equilibrium hole-rich and
hole-poor phases. Since $T_{ps}(p)$ increases as the average doping level $p$ 
decreases, the differences between  high and low local densities are enhanced in low
doping compounds. With the charge structure that comes out of these
calculations in a $N\times N$ cluster, we
use the BdG method, with the local chemical potential associated with
this variable charge structure, to calculate 
the superconducting properties. This is a novel approach to the charge
inhomogeneities in the BdG context. Several interesting features comes
out of this procedure which allows us to study  the insulator, metallic and 
superconducting phases  by 
quantitative calculations that compares well with the experimental
results.


 
\begin{figure}[!ht]
\includegraphics[height=7cm]{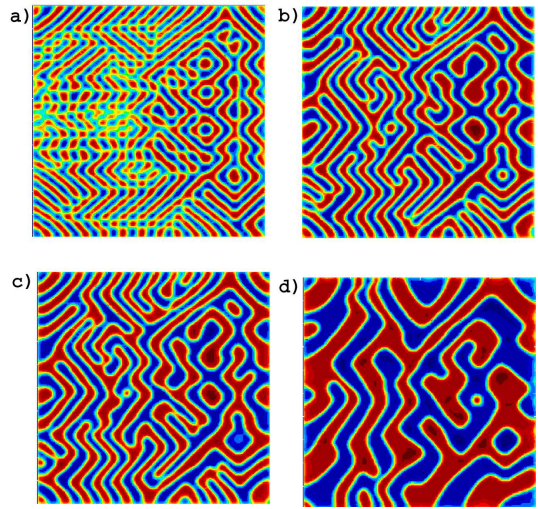}
\caption{ (Color online) The mapping of the order parameter during  the  process of 
phase separation in a compound with average doping  $p=0.12$,
starting with variations
around $u=0$. Panel a) is for t=400 time steps, b) t=800,
c) t=1000 and d) t=4000.}
\label{1801ac}
\end{figure}

The CH non-linear differential equation which describes the time
evolution process of a phase 
separation process, at a temperature $T$ below
the phase separation transition at $T_{ps}(p)$, can be written as a time
($t$) derivative\cite{CH,Mello04}:

\begin{eqnarray}
\frac{\partial u}{\partial t} = -M\nabla^2(\varepsilon^2\nabla^2u
+ A^2(T)u-B^2u^3).
\label{EqCH}
\end{eqnarray}
where $u$ is the local order parameter. In a finite size scheme\cite{Mello04},
it is associated with the local charge $p(i)$, i.e.,
the  local variation from the average number of holes per copper atom $p$, defined at
a site $i$ as $u(i)\equiv p(i)-p$ and 
$u(i) \approx 0$ above and near the $T_{ps}$. $\varepsilon=0.01$ 
and $B=1$ are fixed parameters,  $A$
depends on the temperature $T$ and 
the ratio $\pm A/B$ yields the two local equilibrium densities $p_{\pm}$, 
a low and a high density. 
$M$ is the mobility
of the particles and it dictates the phase separation time scale. Thus,
the degree of
phase separation depends on how fast the system is quenched, what is an explanation
why similar compounds may exhibit different degree of inhomogeneity. 
As the temperature
goes down below $T_{ps}$, the two equilibrium order parameters
(or densities) increases their differences and the energy barrier
between them, $E_g$, also increases\cite{Otton}. $E_g=A^4(T)/B$
which is proportional to $(T_{ps}-T)^2$ and can be taken as the
energy associated with the upper pseudogap temperature\cite{Mello04}.
In Fig.(\ref{1801ac}) we display the mapping of the order parameter for
a $100 \times 100$ system with  $p=1/8$. 
The two equilibrium local densities for this system is
$p_-=zero$ and $p_+=0.24$. 
At the beginning, the phase separation process is quite symmetric,
and could be the reason for the symmetric phases as stripes and checkerboard\cite{Hanaguri}
order.
As it evolves in time, 
the systems tends towards a complete phase separation, with larger
and less ordered stripes, as shown in Fig.(\ref{1801ac}). 

Our main point here is that this phase separation process is
the cause of the charge inhomogeneity found in many HTSC materials. 
Taking the large pseudogap,
as the phase separation temperature $T_{ps}(p)$, it is possible to
infer that a HTSC compound, depending on its doping level, 
evolves in different patterns as, for instance, in the way depicted
in Fig.(\ref{1801ac}). Therefore underdoped compounds which, in general, have
a very high $T_{ps}$, may phase separate into a complete bimodal charge
distribution with larger charge stripes or phase domains (Fig.(\ref{1801ac}c or d)), 
while compounds with $p\approx 0.19$ which have very low values of $T_{ps}$,
segregates in small and symmetric regions as shown in Fig.(\ref{1801ac}a or b). 
Above $T_{ps}$ or for compounds
with $p > 0.19$ the systems are described by small doping variations
around the mean value or a Gaussian distribution. 
This change in  behavior  near $p\approx 0.19$ 
has been detected  in many experiments\cite{Tallon}. 
More recently,  STM data\cite{McElroy1,McElroy2} detected 
the vanishing of the zero temperature pseudogap peaks (ZTPG) for
compounds approaching  $ 0.19$. 
In the CH scenario, the origin of the ZTPG 
is the value of the energy barrier $E_g$  between
the hole-poor and hole-rich phases\cite{Mello04}. Since
the charge domains and $E_g$ decrease with p, 
the ZTPG must behave in a similar fashion:
decreasing in number and intensity and  vanishing near $p=0.19$, 
exactly as observed in recent STM results\cite{McElroy1,McElroy2}.



Now, that we have derived how the 
charge inhomogeneity sets in a HTSC material 
below the high pseudogap temperature $T_{ps}$, we can study 
how the superconductivity develops in a charge
inhomogeneous system, like in a stripe phase, as function of the temperature.
With this aim, we feed into a  local superconducting calculation with the 
BdG mean-field theory\cite{Franz,Ghosal2}, the charge or domains
mappings derived from the CH solutions. As usual, the 
method starts with
the extended Hubbard Hamiltonian

\begin{eqnarray}
&&H=-\sum_{\ll ij\gg \sigma}t_{ij}c_{i\sigma}^\dag c_{j\sigma}
+\sum_{i\sigma}(\mu_i)n_{i\sigma}  \nonumber \\ 
&&+U\sum_{i}n_{i\uparrow}n_{i\downarrow}+{V\over 2}\sum_{\langle ij\rangle \sigma
\sigma^{\prime}}n_{i\sigma}n_{j\sigma^{\prime}},
\label{Hext}
\end{eqnarray}
where $c_{i\sigma}^\dag (c_{i\sigma})$ is the usual fermionic creation (annihilation)
operators at site ${\bf x}_i$,
spin $\sigma \lbrace\uparrow\downarrow\rbrace$, and
$n_{i\sigma} =  c_{i\sigma}^\dag c_{i\sigma}$.
$t_{ij}$ is the  hopping between site $i$ and $j$. In an attempt to
model real systems, we have used
hopping values up to $5^{th}$ neighbors derived from the ARPES data 
of YBCO\cite{Schabel}.
In their notation,
the hopping parameters here are:
$t\equiv t_1$=0.15eV, $t_2$/$t_1$=-0.70,
$t_3$/$t_1$=0.25, $t_4$/$t_1$=0.08, $t_5$/$t_1$=-0.08.
$U=1.1t$ is the on-site and  $V=-1.1t$ is the nearest neighbor
phenomenological interactions.
$\mu_i$ is the local variable chemical potential which reproduces 
the charge inhomogeneity solutions shown in Fig.(\ref{1801ac}), 
according the CH results. The BdG equations are:

\begin{equation}
\begin{pmatrix} K         &      \Delta  \cr\cr
           \Delta^*    &       -K^*
\end{pmatrix}
\begin{pmatrix} u_n({\bf x}_i)      \cr\cr
                v_n({\bf x}_i) 
\end{pmatrix}=E_n
\begin{pmatrix} u_n({\bf x}_i)       \cr\cr
                 v_n({\bf x}_i)
\end{pmatrix}
\label{matrix}
\end{equation}
 
These equations
are solved self-consistently in clusters from $14\times 14$ to
$24\times 24$ sites for the positive eigenvalues $E_n$ or
quasiparticle excitations and the eigenvectors 
or charge amplitudes $u_n({\bf x}_i)$ and $v_n({\bf x}_i)$, together with the pairing 
amplitudes\cite{Franz}
\begin{eqnarray}
\Delta_U({\bf x}_i)&=&-U\sum_{n}u_n({\bf x}_i)v_n^*({\bf x}_i)\tanh{E_n\over 2k_BT} ,
\label{DeltaU}
\end{eqnarray}

\begin{eqnarray}
\Delta_{\delta}({\bf x}_i)&=&-{V\over 2}\sum_n[u_n({\bf x}_i)v_n^*({\bf x}_i+{\bf \delta})
\nonumber \\
&&+v_n^*({\bf x}_i)u_n({\bf x}_i+ 
{\bf \delta})]\tanh{E_n\over 2k_BT} ,
\label{DeltaV}
\end{eqnarray}
and the hole density is given by
\begin{eqnarray}
p({\bf x}_i)=1-2\sum_n[|u_n({\bf x}_i)|^2f_n+|v_n({\bf x}_i)|^2(1-f_n)],
\end{eqnarray}
where $f_n$ is the Fermi function.

The major difference from previous calculations is in
the way that the disorder is taken into account. Ghosal
et al\cite{Ghosal2} used an impurity potential defined by
a random variable between the limits $[-V,V]$. Here, as mentioned above, we 
feed into the initial conditions the non-constant local charge,
as those shown in Fig.(\ref{1801ac}), 
derived from the CH solutions. With this fixed charge mapping in a $N \times N$
cluster, we let the chemical
potential $\mu_i$ change,  self-consistently, until it yields these
initial  charge domains.

\begin{figure}[!ht]
\includegraphics[height=6cm]{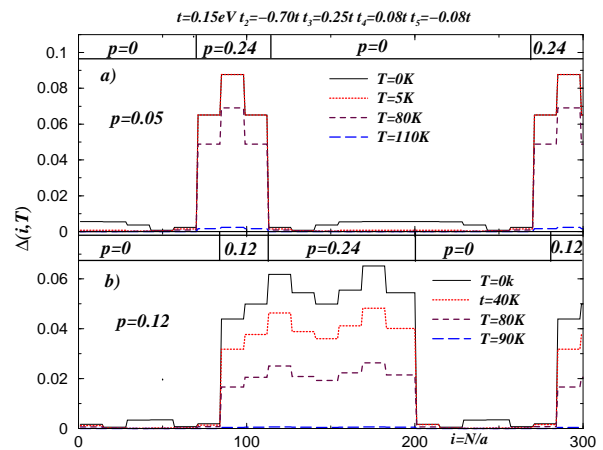}
\caption{ (Color online)
The temperature evolution of the local  gap  $\Delta (i,T)$ (in 
units of $t=0.15eV$)  with stripe charge disorder (the values
at each site {\it i} are given on the top of each panel) and
average doping level $p=0.05$ 
and $0.12$. The horizontal axis displays 21 stripes with
14 unit cells each.}
\label{Deltan012T}
\end{figure}

Fig.(\ref{Deltan012T}) shows the superconducting  d-wave gap 
$\Delta (i,T)$ at each site {\it i} of two $14\times 21$ clusters and their
temperature evolution. Panel a) represents
an underdoped compound of $p=0.05$  with total phase separation
into a bimodal charge distribution made of charge variable stripes of 14 sites each, and
in a less symmetric geometry, as taken from Fig.(\ref{1801ac}d). 
The values of the local doping level $p(i)$ are shown on the
top of each panel. $p=0.0$ is the light color and $p(i)=0.24$
is the dark color phase depicted in the  Fig.(\ref{1801ac}d).
Panel b) represents a compound of average doping of $p=0.12$, made of
values of $p(i)=0.0, 0.12$ and $0.24$ similar to Fig.(\ref{1801ac}c).
The  gaps are calculated by the BdG equations and the chemical potential
$\mu (i)$ evolves self-consistently  at each site $i$, until
it yields the  initial conditions on the local densities, which
are held fixed.
We have verified that changing the stripes configuration in the clusters yields the
same type of gap structure and temperature dependence.

The results for the $p=0.05$ sample with $p_-=0$ and $p_+=0.24$
is shown in the top panel of Fig.(\ref{Deltan012T}). As one can see,
the largest values of the zero temperature gap $\Delta_0(i)$ is in 
the regions where $p(i)=0.24$. This is  a direct interpretation of 
the  1D metallic behavior and the corresponding
high values of the leading edge shift on the Fermi surface\cite{Zhou}
along the ($\pi,0$) and ($0,\pi$) antinodal straight segments.
As $p$ increases, the size of hole-rich stripes  increases, 
for instance, for $p=0.05$ they are typically made of 3 lines   
and for $p=0.12$ they are made of 6 lines. 
Thus, upon doping, 
the hole-rich stripes increases in number and size, changing the 
properties of the system from one
dimensional to two dimensional character and
enhances the overall metallic electronic behavior. 
Such changing with the average doping level was also verified  by 
the ARPES data on many lightly doping samples\cite{Zhou} 
by the measurement of spectral weight 
at the antinodal  and, the increase of the spectral weight 
along the ($\pm\pi,\pm\pi$) nodal regions with $p$.

We discuss now how $T_c(p)$  and $T^*(p)$ (the lower pseudogap) 
are estimated in our calculations.
We see from Fig.(\ref{Deltan012T}) that, at zero temperature, there is
a large variation of $\Delta (i,T)$ throughout the samples. Hole-poor
regions have $\Delta (i,T)\approx 0$ and hole-rich have finite values. As the
temperature increases slightly, mostly local gaps at the hole-poor regions
vanish, leaving only finite superconducting gaps at the hole rich regions. 
Thus the $p=0.05$ sample, as concerns it
resistivity, is an insulator at all finite temperatures, although it has a few
metallic stripes which were detected by ARPES\cite{Zhou}.
For the $p=0.12$ compound, even the gaps at the hole-poor stripes remain  up to
$T=40K$. Thus, it is clear that it is the 
superconducting  percolating temperature
which, in agreement with previous work\cite{Mello03}, 
is the superconducting critical temperature $T_c$. 
In other words, above $T_c$, the local superconducting  gaps
vanish at hole-poor regions  and it  prevents the 
superconducting regions to percolate,  there
are still  some non-vanishing  $\Delta (i,T)$ at the 
hole-rich or metallic regions but, since they occupy marginally less than
$50\%$ of the system size, it cannot hold a
superconducting current. Thus, above $T_c$, the $p=0.12$ sample has a 
metallic behavior. As it is also shown in Fig.(\ref{Deltan012T}), 
these local gaps decreases continuously as the temperature increases further 
and, for  $p=0.05$
and  $p=0.12$, they totally vanish  at $T=110$K and $T=90$K respectively.
This is the interpretation to the onset of the local superconducting or
lower pseudogap temperature $T^*(p)$. 

\begin{figure}[!ht]
\includegraphics[height=6cm]{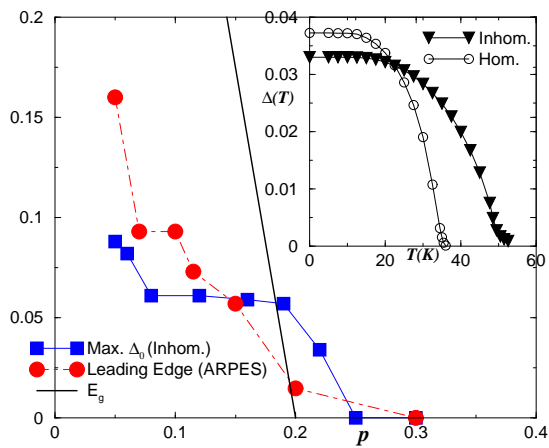}
\caption{ (Color online)
The maximum local zero temperature gap $\Delta_0 (i)$ (squares) compared
with the experimental leading edge shift\cite{Ino} (circles) as function
of $p$. It is also shown the phase separation line or upper pseudogap  energy ($E_g$). 
In the inset  $\Delta(T)$ for the stripe and
homogeneous system. All the gaps and energies are in units
of $t=0.15eV$.}
\label{Leadingedge}
\end{figure}

The  maximum zero temperature $\Delta_0 (i)$ as function of $p$ is shown
in Fig.(\ref{Leadingedge}). $\Delta_0 (i)$
scales reasonable well with the ARPES measurements of the leading edge
shift\cite{Ino} demonstrating that the maximum gap in a sample
increases with the disorder. As already shown in Fig.(\ref{Deltan012T}),
a system with density disorder in the form of stripes, enhances 
the local gap. Thus, we find that the maximum gap for a such 
inhomogeneous sample is larger than that of
a homogeneous similar (same $p$), that is, the disorder
enhances the local gaps at the hole-rich regions and at their
borders. However, with the same coupling, 
the average gap in a disordered system is less than the gap of a homogeneous
one, and  the average
values of $T_c$ is also enhanced. These results are shown, for s-wave calculations, in the
inset of Fig.(\ref{Leadingedge}) and are in agreement with some recent 
calculations\cite{Dagotto}. 

 In summary we have used the CH phase separation approach to model
the inhomogeneity found in LSCO. The derived charge 
stripe-like disorder describes well the dual nature of the
underdoped electronic structure. The energy barrier $E_g$
between the two equilibrium densities  furnishes an interpretation to
the large pseudogap energy scale and the ZTPG measured on Bi2212 by STM. The spinodal
separation provides also a mechanism 
of how the electronic structure of LSCO, as seen by ARPES, 
evolves with doping; the increase
of the leading edge shift at the
straight (1D) segments near the antinodal region and the increase of the
spectral weight near  the nodal region, with the concomitant 
developing of the metallic behavior.  The  disorder favor the clustering of metallic
regions embedded in an insulator matrix at the low doping compounds. The 
site dependent  local superconducting gaps at low and high doping regions
provide also an interpretation to the onset of
superconductivity or lower pseudogap $T^*(p)$, in agreement with the phase
diagram and the nonconventional properties of HTSC.
 
This work has been partially supported by CAPES, CNPq and CNPq-Faperj
Pronex E-26/171.168/2003.


\begin{references}
%
\bibitem{DHS}  A. Damascelli, Z. Hussain, and Z.-X. Shen, Rev. Mod. Phys.
{\bf 75}, 473 (2003)
%
\bibitem{Zhou} X. J. Zhou,  at al,
Phys. Rev. Lett.{\bf 86}, 5578 (2001),
%
\bibitem{Zhou04} X. J. Zhou, at al,
Phys. Rev. Lett.{\bf 92}, 187001 (2004).
%
\bibitem{Ino} A. Ino, et al,
Phys. Rev. B{\bf 65}, 094504,
(2000)
%
\bibitem{Yoshida} T. Yoshida, at al, Phys. Rev.  Lett.{\bf 91}, 027001, (2003).
%
\bibitem{Tranquada} J.M.Tranquada, et al ,
Nature (London),{\bf 375}, 561 (1995).
%
%
\bibitem{Pan}  S. H. Pan et al.,
Nature, {\bf 413}, 282-285 (2001).
%
\bibitem{Davis} K.M. Lang, et al,
Nature, {\bf 415}, 412 (2002).
%
%
\bibitem{Hoffman} J. E. Hoffman, et al,
Science {\bf 297}, 1148-1151 (2002).
%
\bibitem{McElroy1} K. McElroy, et al,
Nature, {\bf 422}, 520 (2003).
%
\bibitem{McElroy2} K. McElroy, et al,
Phys. Rev. Lett., {\bf 94}, 197005 (2005).
%
%
%
%
%
%
%
%
\bibitem{Mourachkine} A. Mourachkine,  Mod. Phys.
Lett.  B, 2005, and condmat-0506732.
%
 \bibitem{Zaanen} J. Zaanen, and O. Gunnarson, Phys. Rev. B {\bf 40}, 7391 (1989).
 %
 \bibitem{Emery} V.J. Emery, and S.A. Kivelson, Nature, {\bf 374}, 434 (1995).
 %
 \bibitem{Grenier} J.C. Grenier, et al,
Phys. C {\bf 202}, 209 (1992).
 %
 \bibitem{Jorg} J. D. Jorgensen, et al,
Phys. Rev. B {\bf 38}, 11337 (1988).
 %
\bibitem{Truccato}  M. Truccato, et al, 
Appl. Phys. Lett. {\bf 86}, 213116, (2005), and condmat-506198.
 %
%
\bibitem{Tallon} J.L. Tallon and J.W. Loram, Physica C {\bf 349}, 53 (2001).
 %
\bibitem{Mello04} E.V.L. de Mello, and E.S. Caixeiro, Phys. Rev. B {\bf 70},
224517  (2004).
%
 %
 \bibitem{CH} J.W. Cahn and J.E. Hilliard, J. Chem. Phys, {\bf 28}, 258
 (1958).
\bibitem{Otton} E.V.L de Mello, and Otton T. Silveira Filho
Physica  A {\bf 347}, 429 (2004).
%
\bibitem{Hanaguri} T. Hanaguri, et al ,
Nature {\bf 430}, 1001 (2004).
%
\bibitem{Franz}
M. Franz, C. Kallin, A.J. Berlinsky, and M.I. Salkola,
Phys. Rev. B {\bf 56}, 7882 (1997).
%
%
\bibitem{Ghosal2}
A. Ghosal, M. Randeria, and N. Trivedi,
Phys. Rev. B {\bf 65}, 014501 (2001).
%
\bibitem{Schabel}
M. C. Schabel, et al, 
Phys. Rev. B {\bf 57},6090 (1998).
%
\bibitem{Mello03}  E.V.L. de Mello, E.S. Caixeiro, and
J.L. Gonz\'alez,  Phys. Rev. B {\bf 67}, 024502 (2003).
%
\bibitem{Dagotto} K. Aryanpour, et al, cond-mat/0507588.
%

\end{references}
\end{document}